\documentclass[runningheads]{cl2emult}

\usepackage{makeidx}  
\usepackage{graphicx} 
\usepackage{subeqnar} 
\usepackage{multicol} 
\usepackage{cropmark} 
\usepackage{eso}      
\makeindex            

\begin{document}
\title*{Cosmological Simulations of X-ray Clusters:
The Quest for Higher Resolution and Essential Physics}
\toctitle{Cosmological Simulations of X-ray Clusters}
%
%
\titlerunning{Cosmological Simulations of X-ray Clusters}           
%
\author{Michael L. Norman}
\authorrunning{Michael L. Norman}
%
%
\institute{UC San Diego, 9500 Gilman Dr, La Jolla, CA 92093-0424}

\maketitle              

\begin{abstract}
I review cosmological simulations of X-ray clusters. 
Simulations have increased in resolution
dramatically and the effects of radiative cooling, star formation
feedback, and chemical enrichment on the ICM are being simulated.
The structure and evolution of non-radiatve X-ray clusters 
is now well characterized. Such models fail to 
reproduce the observed $L_x-T$ relation, implying the need for additional
physics. Simulations adding radiative cooling produce too much cool
gas and unreasonably high X-ray luminosities. Simulations including
star formation and feedback appear more promising, but need further
refinement. New observations should help in this regard.
\end{abstract}

\section{Introduction}
As the largest gravitationally bound objects in the universe, clusters of 
galaxies have attracted the attention of observers and numerical simulators 
alike. For over a decade, beginning the with pioneering hydrodynamic 
simulations of Evrard (1990), 
numerical simulations have been used to understand the physics of X-ray cluster 
formation and to predict their abundance at high redshift which is a sensitive 
probe of cosmology (see review Henry in these proceedings.)
Observationally, clusters of galaxies have historically been studied in 
the optical
and X-ray portions of the EM spectrum (Forman \& Jones 1982). 
X-rays in particular provide 
an unambiguous method for detecting clusters at low and intermediate
redshift, and many surveys have been conducted (Henry, these proceedings.)

A number of groups have simulated the formation
of statistical ensembles of X-ray clusters 
(Kang et al. 1994; Bryan et al. 1994a,b; Bryan \& Norman 1998; 
Eke, Navarro \& Frenk 1998; Yoshikawa, Jing \& Sato 2000) 
in order to provide a theoretical bridge between
what is observed--the X-ray luminosity function (XLF) and X-ray temperature
function (XTF)--and the cluster mass function (CMF). The CMF in turn is
directly related to the matter fluctuation power spectrum $P(k)$--one
of the holy grails of observational cosmology.
In so doing, simulators have discovered that X-ray clusters 
are not the simple gas-bags they were once thought to be. 
It has been found that quite high resolution is required to converge on the
predicted properties of non-radiative clusters (Anninos \& Norman 1996;
Frenk et al. 1999), and that the inclusion
of radiative cooling and other non-adiabatic effects strongly affects the
clusters' emission and structural properties (e.g., Pearce et al. 2000).

This review will follow the development of cosmological simulations of X-ray
clusters primarily from a historical perspective, starting with the 
non-radiative simulations of Evrard (1990) and others and
concluding with current models incorporating cooling, star formation,
supernova feedback and chemical enrichment. 
The field has been enlivened
by the arrival of new observations and new questions. I will attempt to
keep the questions at the forefront of this review, for while some have
been convincingly answered, many are still open. 
Simulations of cluster mergers done outside the framework of CDM-driven 
structure formation are not reviewed here for space reasons.

\section{Methodology}
\begin{figure}
\centering
\includegraphics[width=.6\textwidth]{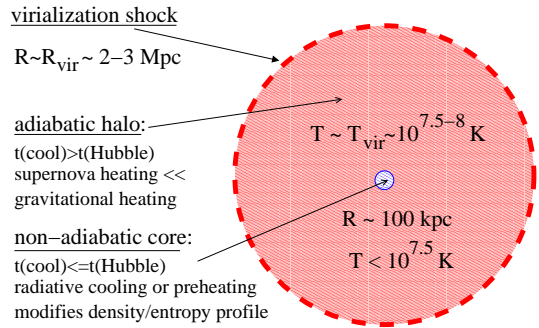}%
\caption{Simplified structure of a bright X-ray cluster.}
\label{cartoon}
\end{figure}
From a purely technical perspective, simulating X-ray clusters has been
a quest for higher resolution and essential physics. This quest is still
ongoing. Here I outline some of the basic requirements for simulations of
this sort, describe how they are performed, 
and how they have advanced over the past decade. 

Fig. 1 shows a simplified model of an X-ray cluster. 
It is a gross generalization
based both on our models of hierarchical structure formation and 
observations of real X-ray clusters. As intergalactic gas falls into the
potential well of the cluster, it passes through several strong shocks,
the innermost of which is called the virialization shock. The virialization
shock is located about one virial radius from the cluster center and it
heats the gas to $T\sim \frac{1}{2}T_{vir}$. Typical numbers for a Coma-like
cluster are $R_{vir}\sim 2-3$ Mpc, $T_{vir}\sim 10^8$ K. 
Inside the virialization
shock is the hot intracluster medium we observe in X-rays. The bulk of the
X-ray halo is adiabatic insofar as 
$t_{cool}=c_v\rho T/\Lambda_x(T)>t_{Hubble}$. Here $\Lambda_x(T)$ is the
bolometric X-ray emissivity of the gas. Inside the adiabatic halo is a
non-adiabatic core where densities are high enough to reverse the inequality.
In the core, radiative cooling
and possibly other processes have modified the entropy of the gas. The
bulk of a cluster's luminosity is emitted from the core region, which is
observed to have a radius of $100-200$ kpc (Sarazin 1986).

Numerical simulations of cluster formation follow the growth of 
density perturbations in the gas and dark matter 
subject to self-gravity and cosmic expansion. The density perturbations
are initialized as a Gaussian random field with a power spectrum 
$P(k)$ taken from linear theory and constrained by observations
(e.g., Tegmark \& Zaldarriaga 2002).
The simulation volume is typically cubic and periodic boundary
conditions are assumed. 
The free parameters in these 
simulations are the usual cosmological ones: $h, \sigma_8, \Omega_d,
\Omega_b, \Omega_{\Lambda}$. Evolving the density perturbations into
the nonlinear regime requires solving the equations of gas dynamics 
and dark matter dynamics subject to their mutual self-gravity in a 
frame comoving with the expanding universe. Both N-body and grid 
methods are used for this purpose (see review by Bertschinger 1998.) 
These simulations must be done in 3D, making them computationally
expensive.

In order to accurately predict the X-ray luminosity of a simulated cluster,
the density and temperature distribution of gas within the core must be 
resolved (Anninos \& Norman 1996). Ten resolution elements (grid points or 
SPH smoothing lengths) in a core radius would be ideal; three would be marginal. 
Here is where the central difficulty arises. A Coma-like cluster forms due to
infall of material within a Lagrangian volume of radius $\sim 15$ Mpc 
comoving. Thus the minimum box size would be 30 Mpc on a side. Taking
30 kpc as the minimum useful resolution, we see that any simulation
will need a spatial dynamic range of 1000 but preferably three times
that. However,
in order to adequately sample the large-scale tidal field which 
affects how the cluster collapses, a box 100 Mpc on a side is
needed. The required spatial dynamic range is thus 3,300 (marginal) or
10,000 (ideal). Simulations aimed at computing cluster statistics (i.e.,
XLF or XTF) would need even larger volumes (250-500 Mpc) and concommitantly 
higher spatial dynamic range.

Simulations employing uniform Eulerian grids cannot achieve such dynamic
ranges because the memory requirement scales as the cube of the number
of grid points. Grids of $1024^3$ are at the limit of the resources of even 
the largest
parallel supercomputers. Instead, researchers employ numerical methods 
with variable and/or adaptive resolution. Two principal methods have been used: 
smoothed particle hydrodynamics (SPH), a gridless particle-based method
that tracks the Lagrangian motion of fluid particles
(Evrard 1990; Thomas \& Couchman 1992; Katz \& White 1993; Metzler \&
Evrard 1994; Navarro, Frenk \& White 1995; Suginohara \& Ostriker 1998;
Lewis et al. 2000; Pearce et al. 2000; Valdarnini 2002)
and a number of grid-based methods employing
static-nested or adaptive Eulerian mesh hierarchies
(Anninos \& Norman 1996; Bryan \& Norman 1997; Loken et al. 2002)
or quasi-Lagrangian deformable meshes
(Gnedin 1995; Pen 1995). 
Of the mesh-based methods, adaptive mesh refinement
(AMR; Bryan 1999; Bryan \& Norman 1999) appears to be the most powerful. 
A comparison of 12 
codes embracing all these methods on the formation of the 
``Santa-Barbara Cluster"--a Coma-like cluster 
in a standard CDM universe--showed that AMR and SPH produce 
comparable results at comparable resolution (Frenk et al. 1999).

Until rather recently, cluster simulations have assumed the gas
is non-radiative because this is the simplest assumption one
can make and it is approximately true. In such simulations, 
shock waves and turbulent mixing of fluid elements are the only 
mechanisms which change the entropy of the gas. Despite this
simplification, convergence on the properties of so-called
adiabatic clusters (a misnomer) has been difficult 
for reasons of resolution and the numerical treatment of shock
waves (Frenk et al. 1999). The effects of turbulent mixing has
scarcely been addressed (but see Norman \& Bryan 1998).
Next in the chain of complexity is the addition of
radiative cooling into the gas energy equation. A number of
authors have carried out simulations including radiative cooling
with interesting albeit divergent results, as discussed
below. Even more ambitious is the inclusion of a recipe for
star formation and feedback (energy and metals) from galaxies.
Simulations of this sort are in their infancy, but show some
promise to recover the properties of real clusters 
(cf. Valdarnini 2002).

\section{Simulations of Non-radiative Clusters}     

\subsection{Can We Make Coma?}
Evrard (1990) carried out the first hydrodynamic cosmological simulations
of X-ray cluster formation in a standard CDM model. His motivations were 
twofold: first, to study how a rich cluster of galaxies is assembled in 
hierarchical models, and second, to see if the resulting cluster resembled 
Coma. The code employed (Evrard 1988) coupled a variable smoothing length 
implementation of SPH to a P$^3$M N-body solver for the dark matter.
With only 4096 particles ($32^3$) each for gas and dark matter fields, 
the spatial resolution in the cluster core was 200 kpc. 
While insufficient to resolve the cluster core, the simulations
established a number of results that have been confirmed by subsequent,
higher resolution simulations. First, that the cluster forms at the intersection
of filaments of dark matter and gas by the mergers and accretion of subsclusters. 
Second, that
inside the virial radius, the gas is {\em approximately} isothermal and in 
hydrostatic equilibrium, thus justifying two key assumptions made in
the isothermal $\beta$-model introduced by Cavaliere \& Fusco-Femiano 
(1976) to
fit X-ray cluster brightness profiles. A third interesting result was
that residual kinetic energy in the gas and anisotropy in the dark
matter velocity distribution function biases cluster mass estimates made
using the $\beta$ model on the low side by 30\%. Subsequent studies have shown
that the magnitude of this bias is sensitive to numerical resolution
and the dynamical state or the cluster (Evrard, Metzler \& Navarro
1996). Finally, that the integrated 
cluster properties and synthetic X-ray maps were encouragingly similar to 
Coma and A2256, suggesting that models of this kind were on the right
track.

Thomas \& Couchman (1992), with their own implementation of SPH+P$^3$M
achieved an impressive 20 kpc resolution in the center of the cluster
with $2 \times 32^3$ particles by drastically dropping the
force softening length and SPH minimum smoothing length. This they
did at the risk of introducing spurious two-body relaxation effects
which they analyzed in detail. Confident that the
effect was not present, they found that the gas density profile formed
a flattened central core, while the DM profile possessed a central cusp.
This latter finding foreshadowed the discovery a universal dark matter
density profile in CDM simulations by Navarro, Frenk \& White (1997),
but was not focussed on. Rather, the authors commented that
``the outer profiles match very well". 

\subsection{Cluster Scaling Laws}
Kaiser (1986) showed that in the absense 
of non-gravitational heating/cooling, 
the ICMs of X-ray clusters of different masses
and formation epochs should be self-similar. Kaiser derived scaling laws
relating the mass, radius (and hence density), temperature, 
and DM velocity dispersion for a top-hat
density perturbation collapsing and virializing at redshift z.
Namely: $T(M,z)\propto M^{2/3}(1+z)$ and 
$L_x(M,z)\propto M^{4/3}(1+z)^{7/2}$.
A direct consequence of these scaling laws is that at fixed
redshift, $L_x$ should scale as $T^2$. Observations, however,
are consistent with $L_x \propto T^3$ with substantial scatter 
(David et al. 1993).
Since simulations showed that clusters form
hierarchically, it is reasonable to ask whether Kaiser's scaling 
laws are still obeyed.

With this motivation, Navarro, Frenk \& White (1995) 
simulated the formation of 6 clusters of different masses
drawn from a single, large box standard CDM simulation. 
Their procedure was
innovative: the initial data for each cluster was resampled at
higher resolution and then evolved within the large-scale tidal
field. The simulations were performed with SPH coupled to an N-body
tree code with a force softening length of 100 kpc. 
This study yielded several important results. First, that the
density profiles for both gas and dark matter when plotted as
overdensities versus radius in units of the virial radius exhibited
self-similar behavior. Second, that the gas and dark matter
profiles showed no evidence of a central core, but
rather exhibited a logarithmc slope that steepens from
$\sim -1$ near the center to $-3$ near the edge. They introduced a
fitting function for the dark matter profile whose general
form we now refer to as an NFW profile:
\[\rho(r)/\rho_{crit}=\frac{\delta_c}{(r/r_s)(1+r/r_s)^2}\]
where $r_s$ is a scale radius and $\delta_c$ is a characteristic
density.
They studied the formation history of their six clusters and
found that while merger histories differed considerably from
cluster to cluster, all clusters gained the majority of their mass
between z=1 and z=0.2. Finally, they confirmed that their 
clusters obeyed Kaiser's scaling laws and concluded that 
the central properties of the intracluster medium are 
determined by non-gravitational processes such as radiative
cooling or pre-heating at high redshift.

\subsection{Numerical Resolution and the $L_x-T$ Relation}
In 1994 several papers appeared (Kang et al. 1994; Bryan et al. 1994a,b)
which attempted to compute large statistical samples of non-radiative
X-ray
clusters using Eulerian grid codes combining two higher order-accurate gas
dynamics algorithms (TVD, PPM) with a particle-mesh dark matter
solver (Ryu et al. 1993; Bryan et al. 1995). 
The advantage of these methods relative
to SPH are their speed, allowing larger grids and particle counts
for the same computer resource, and superior shock-capturing
ability. A disadvantage relative to SPH is that the spatial
resolution is limited by the grid spacing $\Delta x$, which is
fixed. As discussed above, achieving even 30 kpc resolution in 
cluster cores is not feasible with current computers without
resorting to nested or adaptive grids.
Nonetheless, Bryan et al. (1994b) simulated
a large sample of X-ray clusters in a cold+hot dark matter (CHDM) 
universe at low resolution and found that the simulated clusters 
obeyed $L_x \propto T^3$, in agreement with observations but at
odds with Kaiser's scaling laws.
Was this a result of new physics (massive neutrinos) or numerical
resolution, as suggested by Navarro et al. (1995)? 

Anninos \& Norman
(1996) carried out a resolution study of non-radiative X-ray
clusters using a nested-grid Eulerian cosmological hydro code
and found that while
the mass-weighted cluster temperature was a weak function of
resolution, the X-ray luminosity was very sensitive to resolution, scaling
as $L_x \propto \Delta x^{-1.1}$ for $1 \leq \Delta x (Mpc) \leq 0.1$.
This dependence is a consequence of under-estimating the central
gas density due to resolution effects and thereby under-estimating the
X-ray emissivity in the cluster core.
Bryan \& Norman (1998) used this scaling law to correct the 
predicted luminosities for three statistical samples in three
cosmologies (SCDM, CHDM, OCDM) computed at
300 kpc resolution and found good agreement with Kaiser's (1986)
scaling laws and the results of Navarro, Frenk \& White (1995). 
The discrepancy in the predicted $L_x-T$ relation
was thus explained as an artifact of poor resolution.
However, although all simulations now agreed, they still disagreed
with observations.  
  
 \subsection{The Santa Barbara Cluster Test Project}
\begin{figure}
\centering
\includegraphics[width=0.3\textwidth]{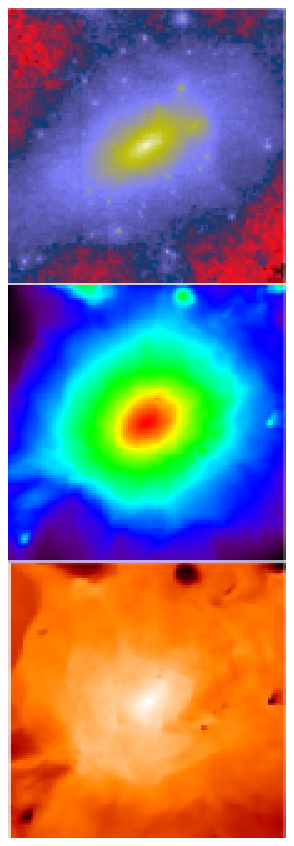}%
\includegraphics[width=0.7\textwidth]{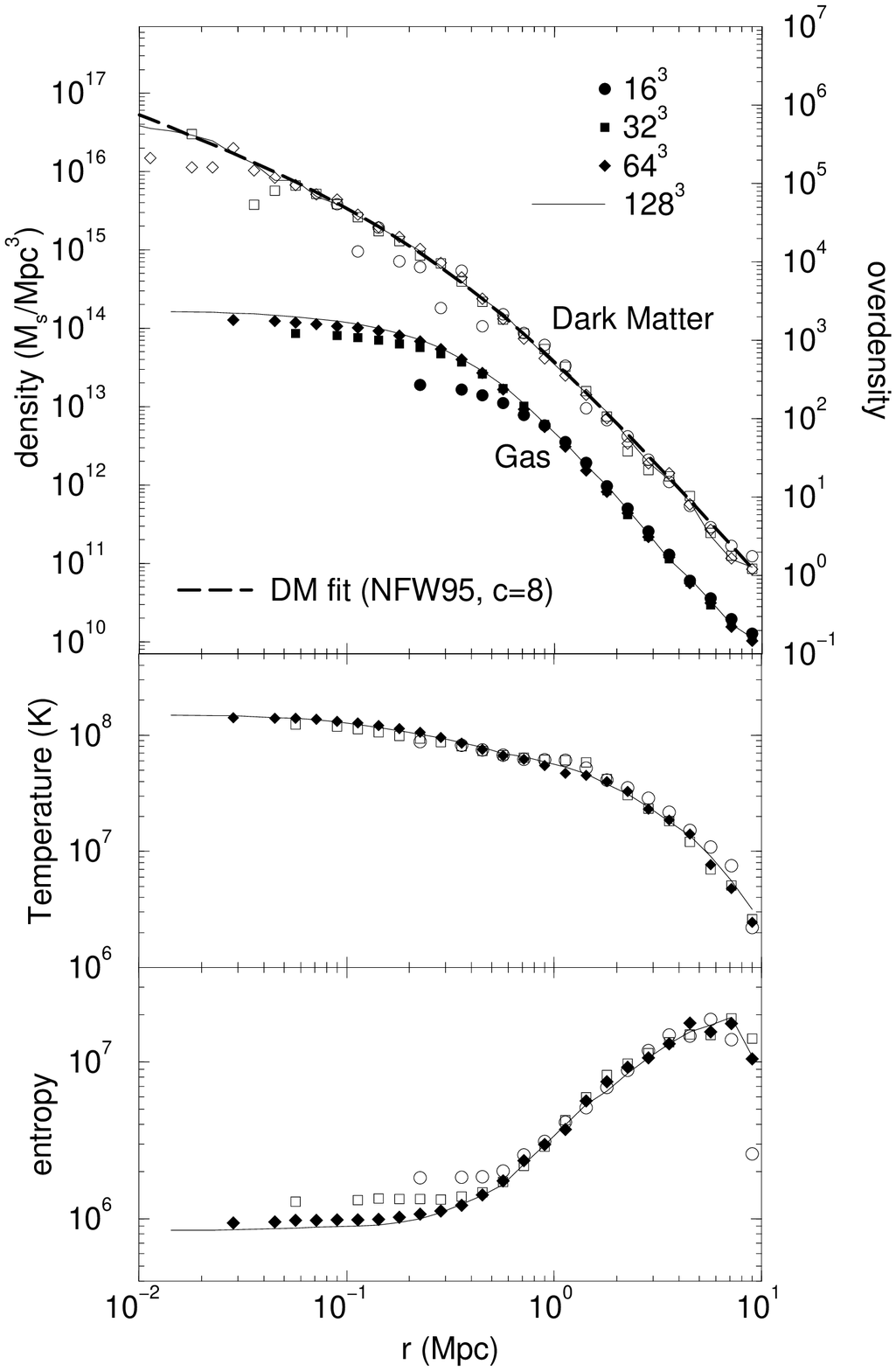}
\caption{AMR simulation of the Santa Barbara cluster.
{\em Left, from top to bottom:} projections of dark matter density,
gas density, and emission-weighted temperature in a 5 Mpc field
centered on the cluster. {\em Right:} 
Radial profiles of spherically averaged quantities, from Bryan \&
Norman (1997).}
\label{profiles}
\end{figure}
By the late 1990s, a number of groups had developed codes capable
of resolving the core structure of non-radiative clusters. The
structure of non-radiative clusters is now fairly well determined.
The definitive study was the Santa Barbara Cluster 
Comparison Project (Frenk et al. 1999) in which the results of
12 codes implementing seven different numerical algorithms were
compared for Coma-like cluster forming in a standard CDM cosmology
from identical initial conditions. Among codes capable of resolving
the core radius of the gas distribution, X-ray luminosities
agreed to within a factor of 2. All codes regardless of
resolution agreed to within 10\% on the mass and average temperature 
of the cluster. This implies that simulations can predict the
XTF to much higher precision than the XLF, ignoring the effects
of other physics.

Fig. 2a shows an AMR simulation of the Santa Barbara cluster
with a spatial dynamic range of 8,192 (Bryan \& Norman 1997).
This corresponds to a 
resolution of 7.8 kpc inside the cluster core within a 64 Mpc
cube. Note that the dark matter halo is
elongated while the gas distribution is more spherical due to the
isotropizing effect of gas pressure. The virialization shock can be
seen as a discontinuity near the edge of the temperature image.
Fig. 2b shows spherically
averaged radial profiles of various quantities from the same
simulation.
These results were included in the Santa Barbara comparison
and are representative of the high resolution results.
The dark matter profile is well fit by an NFW profile with
concentration parameter c=8. At high
resolution, the gas density profile tracks the NFW
profile at large radii but has a flat core at small
radii $r<0.1 r_{vir}$.  The entropy-related variable
$T/\rho^{2/3}$ also
exhibits a flat core, but rises linearly with radius 
outside the core. The temperature profile exhibits a substantial
gradient at large radii and continues to rise, albeit
more slowly, as $r\rightarrow 0$.

\subsection{Cluster Temperature Profiles}
\begin{figure}
\centering
\includegraphics[width=.5\textwidth]{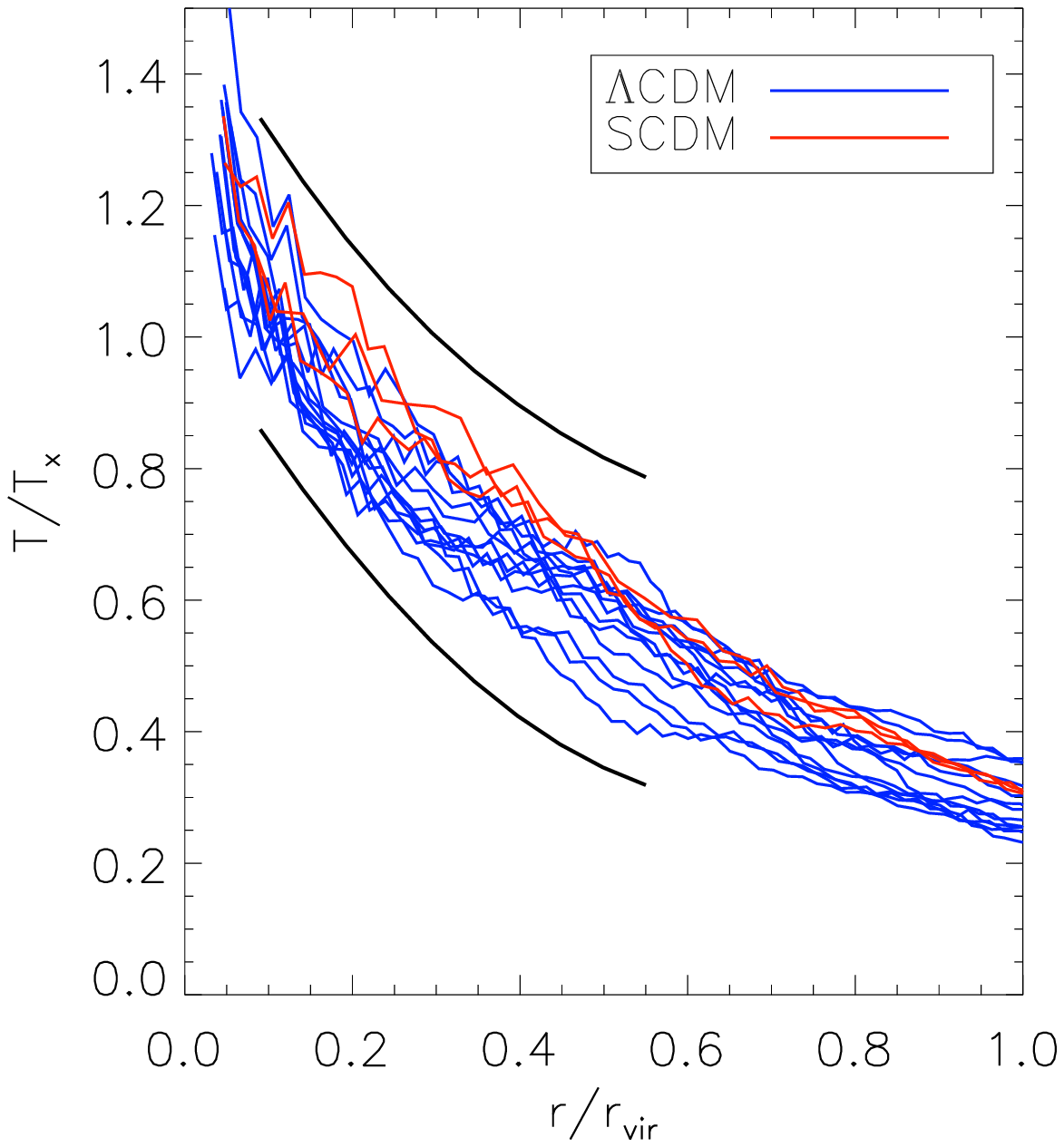}%
\includegraphics[width=.5\textwidth]{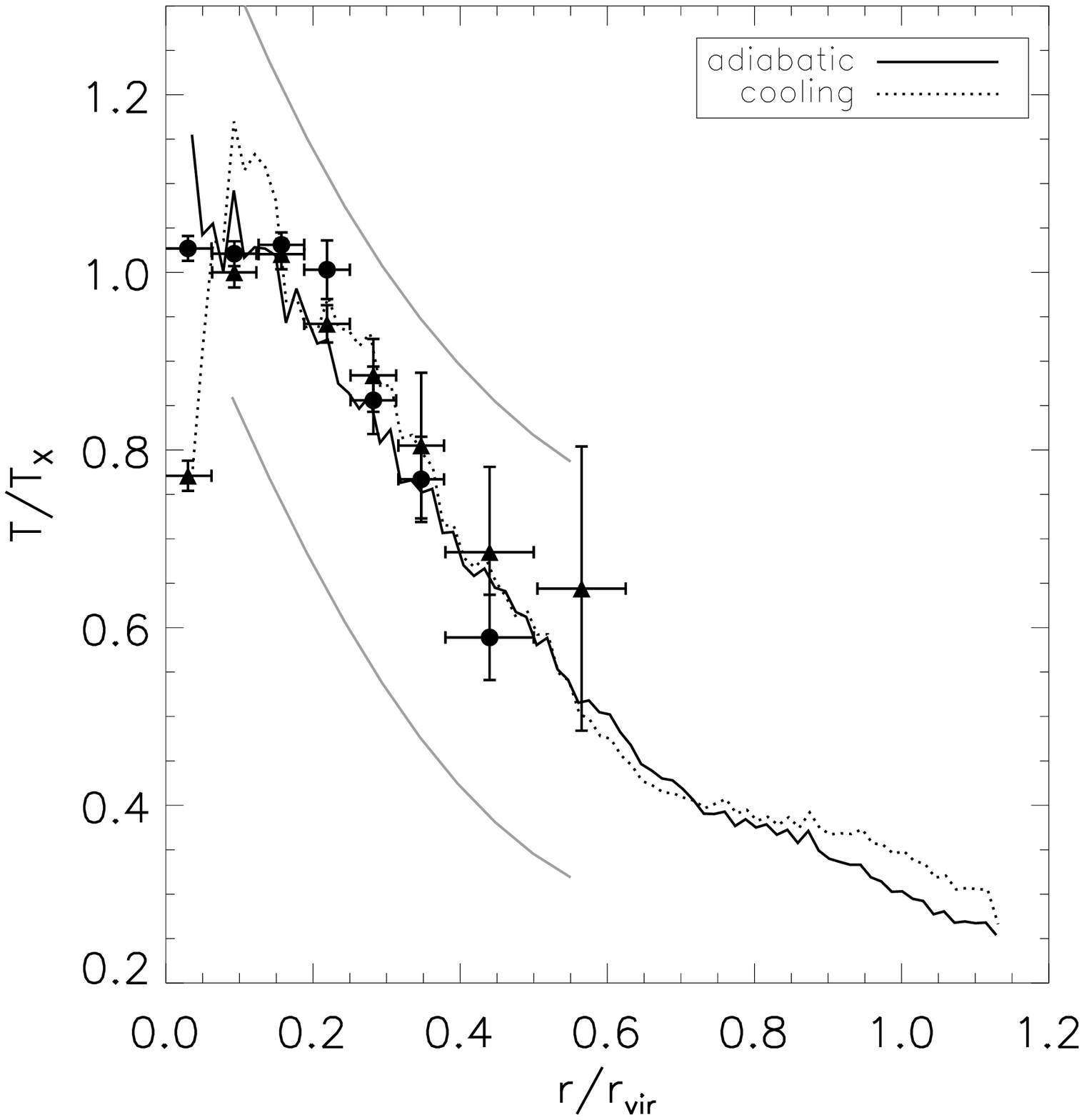}
\caption{(a) Universal temperature profile from AMR simulations
non-radiative, relaxed clusters in two cosmologies,
(b) Comparison of numerical predictions with 
recent $BeppoSAX$ data (De Grandi \& Molendi 2002). In both figures, 
the black solid lines are the $1 \sigma$
confidence band from Markevitch et al. (1998). From Loken et al.
(2002).}
\label{cartoon}
\end{figure}
Early observations and numerical simulations suggested
that ICMs in X-ray clusters are isothermal. Indeed, isothermality
is generally assumed when fitting $\beta$-models to X-ray
surface brightness profiles. This would be a surprising result, if
true, since in gravitationally-bound systems the 
temperature usually rises toward the center. It would 
imply that some mechanism such as thermal conduction was
efficient in transporting energy outward and depositing
it at large radii. Radiation cannot do the job as the 
plasma is optically thin to X-rays. Analyses have shown that
thermal conduction is incapable of rendering a cluster 
isothermal even if the full classical electron conductivity 
is assumed (e.g., Loeb 2002). 

The non-radiative simulations of Navarro, Frenk \& White (1995)
found that the temperature profile is nearly isothermal in
the region that emits most of the X-rays $r<0.4 r_{vir}$, 
but then drops to about one-half the central value at the
virial radius. Results of other SPH codes in the Santa Barbara
cluster comparison project generally agreed, however with
substantial differences in the central value. Some SPH codes found
a central dip in temperature while others did not. 
High resolution grid-based codes, such as AMR, agree well with the SPH
results beyond $r=0.4 r_{vir}$, but continue to rise to small
radius (Fig. 2). Thus, the issue of core temperature profiles
is not yet settled, even for non-radiative simulations.
 
Observationally, the issue is also clouded. 
Markevitch et al. (1998) found evidence of decreasing temperature
profiles in a sample of nearby hot clusters ($>3.5$ keV) observed
with ASCA. 
A subsample of 17 regular/symmetric clusters displayed
remarkably similar temperature profiles
(when normalized and scaled by the virial radius) consistent with
$T\propto [1+(r/r_c)^2]^{-3 \beta (\gamma-1)/2}$ where 
$\gamma=1.24^{+.20}_{-.12}$ and $\beta=2/3$. The typical decrease is 
therefore a factor
of $\sim$2 in going from 1 to 6 core radii (or .09 to 0.5 virial radii).
This result remains controversial as three subsequent studies of large 
samples of clusters concluded that
the majority of cluster temperature profiles show little, or no,
decrease with radius (Irwin, Bregman, \& Evrard 1999; White 2000; Irwin
\& Bregman 2000). Most recently, De Grandi \& Molendi (2002) have presented
a composite temperature profile based on {\it Beppo}SAX data which exhibits an
isothermal core and then decreases quickly.

Loken et al. (2002) computed a statistical sample of non-radiative
X-ray clusters in two cosmologies (LCDM and SCDM) using AMR.
The temperature profiles are well fit by 
$T/T_o=1.3  [ 1 + 1.5 r/r_{vir} ] ^{-1.6}$. This fit is in excellent
agreement with Markevitch et al. (1998) and also
in good agreement with the {\it Beppo}SAX data
outside $r=0.2 r_{vir}$. The simulation results and comparison with
data are shown in Fig. 3. These results suggest
that relaxed X-ray clusters should have a universal temperature
profile outside the core, inside which additional heating and cooling
processes may operate.

\section{Simulations Including Preheating and Radiative Cooling}
\subsection{Preheating}
Give the failure of non-radiative simulations to reproduce the observed
$L_x-T$ relation, one is driven to consider the effect of other physics.
One possibility is an early epoch of preheating 
(David, Jones \& Forman 1991; Evrard \& Henry 1991; Kaiser 1991; White 1991)
that raises the entropy of the intergalactic medium prior to
it being incorporated into a cluster. The idea is that preheating
introduces an entropy floor that breaks the self-similarity between
dark matter and ICM on difference mass scales. If this entropy floor
were larger than the core entropy produced by gravitational heating
in low mass clusters but not so in high mass clusters, 
the central densities and
hence luminosities of low mass clusters would be reduced relative
to high mass clusters, thus steepening the $L_x-T$ relation.
This has been confirmed numerically by Navarro et al. (1995) and 
Pierre et al. (1999). An extensive recent study by
Bialek, Evrard \& Mohr (2001) systematically varied the initial
entropy $S_i$ in a grid of simulations and found that the $L_x-T$
relation, as well as the X-ray size-temperature and mass-temperature
relations steepen monotonically with $S_i$. They found that the 
three observed relations could be satisfied with $S_i \in 55-150$
keV cm$^{-2}$. These levels compare favorably to observational 
determinations of core ICM entropy by Lloyd-Davies et al. (2000).

\subsection{Radiative Cooling, Cooling Flows, and Cooling Catastrophes}
%
The potential impact of radiative cooling by the ICM has been 
recognized for many years (Cowie \& Binney 1977, Fabian \& Nulsen 1977).
The X-ray emitting gas is trapped in approximate hydrostatic
equilibrium in the cluster's potential well. A ``cooling flow" is
believed to form as the radiating gas loses pressure support and
flows inward to higher density values thus accelerating the
cooling rate. In cooling flow clusters, typically, the gas
within approximately 100 kpc of the center has a cooling time
which is less than the age of the cluster. This theoretical 
scenario has been extensively developed (see Fabian
1994 for a review) and observations of X-ray clusters with
central brightness excesses are routinely analyzed with the 
framework of these models (e.g., David et al. 2001).

In principle, cosmological hydrodynamic simulations including 
radiative cooling should see these cooling flows develop 
self-consistently. This effort has not been entirely successful
and the implications of this have not yet been sorted out. Thomas \&
Couchman (1992) and Katz \& White (1993) did the first SPH simulations
with cooling and found that an unresolved lump of cold ($~10^4$ K)
gas forms in the cluster center. In the simulation of Katz \& White
the lump contained $30\%$ of the baryons causing the integrated
luminosity to be too high for a cluster of that temperature.
Similar results have been obtained by Suginohara
\& Ostriker (1998) and Valdarnini (2002)
using higher resolution SPH simulations. 
While superficially resembling a cooling
flow cluster, it is clear that in the absense of star formation
and feedback, too much gas cools. Solutions to the so-called overcooling
problem have been proposed, including conversion of gas to stars 
(Valdarnini 2002), feedback from star formation, and 
heating from a central AGN.

Pearce et al. (2000) discussed a purely numerical origin of the
overcooling problem due to the way SPH treats phase boundaries
(see also Springel \& Hernquist 2002). 
If one imagines how an ideal discontinuity between a cool, dense
phase and a hot diffuse phase is represented in SPH, one realized
that the kernel averaging will smear the discontinuity over several 
smoothing lengths and thereby create a region of intermediate
density and temperature of that thickness. Since the X-ray 
emissivity is proportional to $\rho^2 T^{1/2}$, the numerically-induced 
layer's emissivity will be boosted and
will radiate energy at a rate which is proportional to the
thickness of the layer. In reality, the layer would be a few
collision mean free paths thick but numerically is it much larger. 
The enhanced cooling thereby leads to too much hot gas being
converted to cool gas. 

In order to overcome this problem,
Pearce at al.~(2000) carried out simulations
of cooling clusters with a formulation that manually limits
how much gas may cool. Using values from observed clusters,
they found that cooling had a {\em global effect} on the cluster
profiles. Namely, that cooling produces an inflow of high entropy
gas from the outer parts of the cluster, raising the cluster temperature
and decreasing the X-ray luminosity. Outside the cooling radius, 
the temperature profiles were found to be in good agreement
with Markevitch et al. (1998); while inside the cooling radius, the
temperature decreased toward the center as seen in cooling flow
clusters.

\begin{figure}
\centering
\includegraphics[width=.4\textwidth]{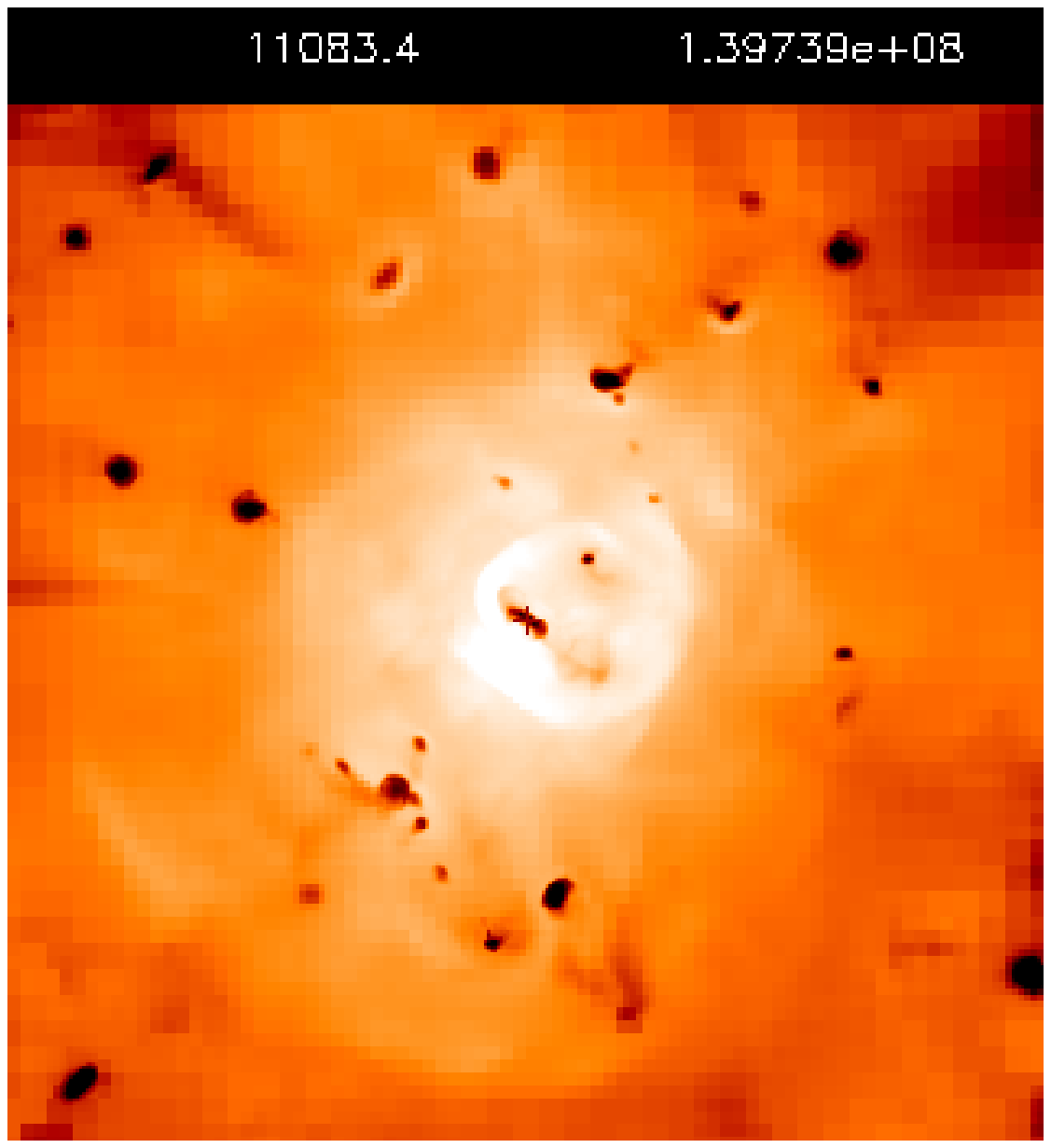}%
\includegraphics[width=.4\textwidth]{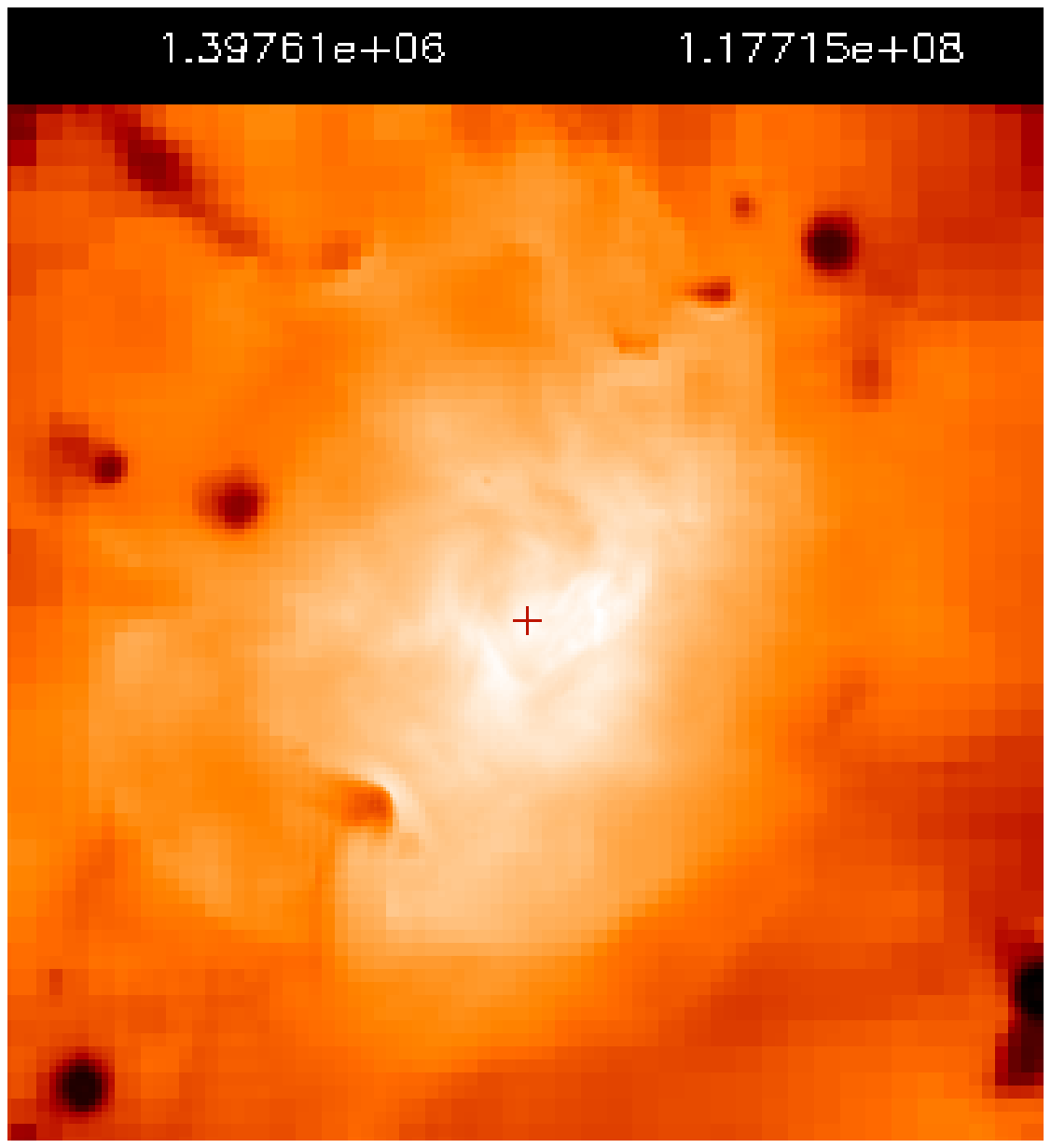}
\caption{Emission-weighted temperature image for a cluster
simulated (a) with, and (b) without radiative cooling included.
The image is 5 Mpc on a side.}
\label{cartoon}
\end{figure}

Does the presence of cool gas in clusters cores necessarily 
imply the existence
of cooling flows? Motl et al. (2002) studied this question with
high-resolution hydrodynamic AMR simulations and concluded no. 
Motl et al. computed two massive X-ray clusters in a $\Lambda$CDM
universe with and without radiative cooling included. The simulations
achieved a spatial resolution of 22 kpc in a cosmological
volume 365 Mpc on a side. Fig. 4 shows a comparison of the non-radiative
and radiative simulations. The cooling simulation produces 
lumps of cooler gas that are associated with subclusters. These
subclusters merge hierarchically and survive their passage through the
ICM.  Those on radial orbits find their way to the cluster core
where they deposit their cool gas. Hierarchical formation 
naturally explains the high frequency of cooling cores in rich galaxy 
clusters despite the fact that a majority of these cluster show evidence 
of substructure which is believed to arise from recent merger activity.
The simuations also produce cooling fronts, ``bullets", and filaments
(Markevitch, these proceedings.)

\section{Simulations Including Galaxy Formation and Feedback}

Although galaxies comprise a small fraction ($15-25\%$) of the baryons
in a rich cluster of galaxies (David et al. 1990), they may have a
disproportionate effect on the ICM. Observations indicate ICM metallicities
of $Z\sim 0.3Z_{\odot}$ (e.g., Edge \& Stewart 1991). These metals 
imply supernova enrichment of the gas either before, during, or after
cluster assembly. Mushotzky \& Lowenstein (1997) report little 
evolution in cluster metallicities out to $z\sim 0.3$, suggesting
that the gas was enriched prior to cluster assembly. Supernova 
enrichment would naturally be accompanied by supernova heating. A
key question therefore is whether this heating is sufficient to 
provide the entropy floor observed in clusters (Lloyd-Davies
et al. 2000) as has been suggested by Ponman et al. (1999).

\subsection{Effects on Structure}

Numerical simulations of cluster formation including galaxy 
formation and feedback are in their infancy. Nonetheless, they
show signs of producing clusters that are more in line with
observations and eliminating the overcooling problem encountered
in cooling--only simulations (Valdarnini 2002). The first study of 
this sort was by Metzler \& Evrard (1994).
Because galaxies form on scales that are unresolved in standard
cluster simulations, special care must be made in how galaxies
are represented. Metzler \& Evrard (1994) put the galaxies in
``by hand" at locations corresponding to peaks in the initial
density field that would eventually form $L_*$ galaxies. A galaxy
particle was introduced at each of these locations in the initial
data and evolved dynamically along with the gas and dark matter 
particles. A total of 108 galaxy particles fed back energy and 
metals according to a user-supplied history. These authors
compared evolutions for a cluster with ``extreme feedback" and
without. Extreme feedback consisted of assuming a constant
feedback level for $0\leq z \leq 4$ and a wind luminosity
of $4 \times 10^{42}$ erg/s for a $10^{10} L_{\odot}$ galaxy.
They found that feedback raised the entropy of the ICM, establishing
an entropy floor in the core. This reduced the gas density in the core
and hence the integrated luminosity. The emission weighted temperature
increased only 15\%, and the X-ray morphology of the cluster was
relatively unchanged.

Lewis et al. (2000) carried out TREESPH simulations of the
formation of a Virgo-mass cluster in a standard CDM cosmology
including star formation and energy feedback. The simulations
convert localized regions of gas which are cooling and collapsing 
rapidly into collisionless star particles according to the 
recipe described in Katz, Weinberg \& Hernquist (1996). In this
approach, an SPH particle is eligible to form stars if it 
exceeds a threshold density and overdensity, and if the
particle's neighborhood is collapsing and Jeans unstable. 
If these conditions are met, 
gas is turned into stars at a rate $d$ ln $\rho_g/dt=-c_*/t_g$, 
where $t_g$ is the maximum of the local dynamical time and
cooling time. The advantage of this approach is that in principle
the star formation histories of individual galaxies can be
computed, as opposed to the constant rate assumption of Metzler \&
Evrard (1994). The disadvantage is that quite high mass
and spatial resolution are required, and it is unclear whether 
simulations have achieved the necessary resolution for convergence.

Lewis et al. report on simulations with and without galaxy feedback
with a spatial resolution of $14 h^{-1}$ kpc. They find that
inclusion of cooling and star formation affects the structure
of the entire cluster, in agreement with Pearce et al. (2000). 
$30 \%$ of the baryons are converted to
stars forming a massive galaxy at the center of the cluster,
altering the cluster potential well. With the low entropy gas
thus converted into a pressureless component, the higher
entropy gas settles into the core region. Compared to the
non-radiative model, they find this structural readjustment
leads to a $20\%$ higher emission-weighted temperature and 
a $30\%$ higher X-ray luminosity. This result is partially
in conflict with the results of Pearce et al. (2000), who
found an increase in $<T_x>$ and a {\em decrease} in $L_x$.
Interestingly, the concentration of baryons in the core was
found to actually steepen the dark matter profile relative to
an NFW profile inside $\sim 30$ kpc.

Lewis et al. found that star formation and feedback 
{\em steepens} the ICM temperature profile in the inner
regions, a result confirmed by Valdarnini (2002).
However, the latter author found that cooling creates
a central temperature inversion within $\sim 50-100$ kpc,
which he interpreted as a cooling flow.
Recent $BeppoSAX$ observations of cooling flow
cluster temperature 
profiles (De Grandi \& Molendi 2002; cf. Fig. 3b) show
this central dip in what is otherwise an isothermal
core extending to $\sim 0.2 r_{vir} \sim 400$ kpc, in
rough agreement with the simulations.

\subsection{Metal Enrichment}

\begin{figure}
\centering
\includegraphics[width=.4\textwidth]{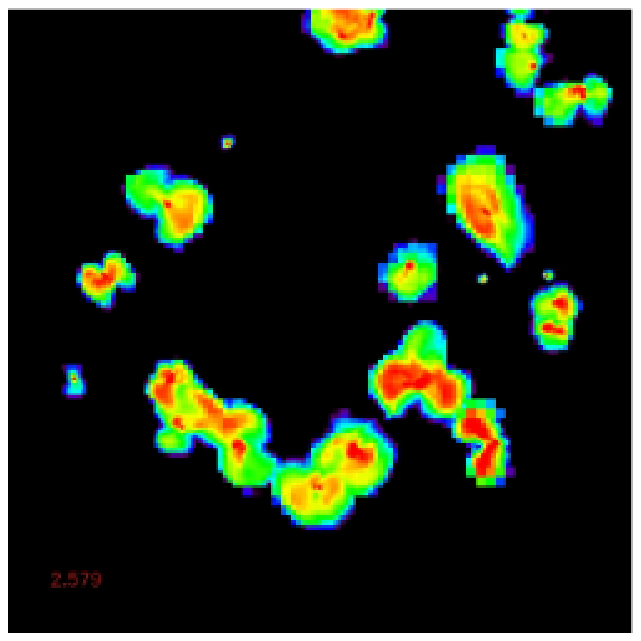}%
\includegraphics[width=.4\textwidth]{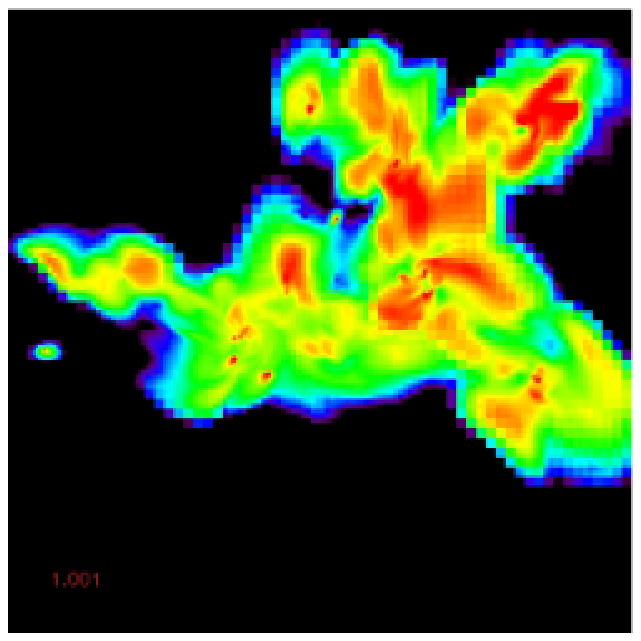}
\includegraphics[width=.4\textwidth]{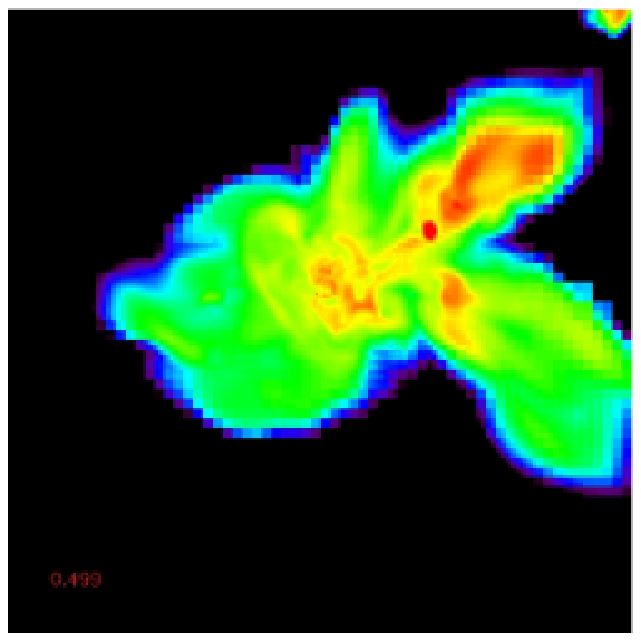}%
\includegraphics[width=.4\textwidth]{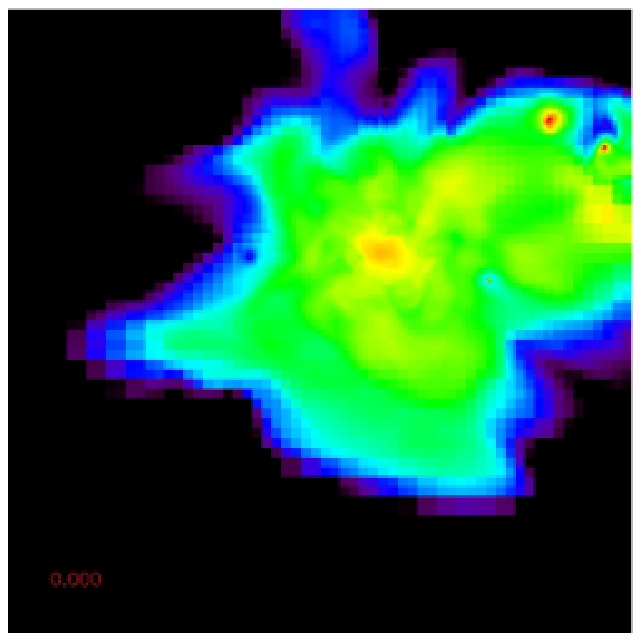}
\caption{Evolution of metallicity field in an AMR simulation
of the Santa Barbara cluster including cooling, star formation,
and feedback. {\em upper left to lower right:} $z=2.6, 1, 0.5, 0.$}
\label{metals}
\end{figure}

If a cluster formed out uniformly mixed intergalactic material, 
and if no subsequent injection of metals took place after the
cluster virialized, then one would expect uniform ICM 
metallicity profiles. If, on the other hand, the infalling gas was 
was enriched locally by the galaxies which ultimately end up
in the cluster, then one would expect a strong negative radial gradient.
If enrichment occured after cluster assembly, then the high density
of galaxies in the inner parts of the cluster would accentuate any
pre-existing metallicity gradient. The same could be said for
cooling flows. Thus, metallicity gradients in clusters can in principle
probe the star formation and chemical enrichment history of a large
sample of the universe (Mushotzky \& Lowenstein 1997).

Metal enrichment was included in the simulations of Metzler \& Evrard 
(1994). In their ``extreme feedback" model, they assumed a constant,
galaxy mass normalized, metal injection rate from z=4 to 0. The transport
of metals from the galaxy particle occurs by virtue of the SPH smoothing
operation rather than resolved hydrodynamically. To the extent that
the smoothing length exceeds/underestimates the physical mixing scale,
the resulting metallicity distribution will be smoother/less smooth
that nature would produce. A simulation with a minimum smoothing
length of 110 kpc produced a metallicity distribution which is
strongly peaked about the cluster center. In units of the wind
metallicity--a free parameter--the metallicity varied from
0.5 in the center to 0.1 at 1 Mpc. Steeper gradients could be achieved 
by reducing the amount of mixing. 

The simulations of Valdarnini (2002) included chemical enrichment
from galaxies formed ``self-consistently" in the protocluster. 
The z=0 cluster exhibited a metallicity distribution that
varied from solar at the center to $<0.1$ solar at $r=100$ kpc. 
This is much steeper that what is observed 
(De Grandi \& Molendi 2001). A likely explanation
for this disagreement is that the galaxy population is severely 
undersampled such that the chemical enrichment is dominated
by the massive central galaxy.

Fig. 4 shows the results of an AMR simulation including cooling,
star formation, feedback and chemical enrichment which produces
a metallicity distribution more in line with observations 
(O'Shea et al. 2002). The Santa Barabara cluster was rerun
with the star formation recipe of Cen \& Ostriker (1993).
The minimum spatial resolution is $15.6$ kpc, which is
sufficient to resolve the formation of a dozen
massive galaxies that enrich the IGM prior to cluster
collapse.  
Metallicity is tracked as a separate fluid dynamic field,
and therefore mixing occurs hydrodynamically. Fig. 4 shows 
the evolution of the metallicity field. At high redshifts,
a population of $\sim 20$ galaxies enriches the IGM over
a region $\sim 8$ Mpc in diameter. The metallicity
distributiuon is extremely inhomogenous at this time. As
time goes on, this gas is drawn into the cluster and mixed
with pristine gas. The final metallicity distribution exhibits
a mild gradient, dropping from $0.3 Z_{\odot}$ at the 
center to $0.1 Z_{\odot}$ at $r=0.3 r_{vir}$. This gradient
is in good agreement with the cooling flow cluster sample of
De Grandi \& Molendi (2001) although the normalization is
too low by a factor of 2. 

\noindent
\flushleft
Acknowledgements: I wish to thank my students and collaborators for 
their contributions to the work described here: Greg Bryan,
Jack Burns, Chris Loken, Patrick Motl, Erik Nelson and Brian O'Shea.


\clearpage
\addcontentsline{toc}{section}{Index}
\flushbottom
\printindex

\end{document}